# Temperature Dependence of the Optical Transition Characteristics of MAPbCl$_x$Br$_{3-x}$ Single Crystals

*Daeyoung Park[2], Yong Ho Shin[1], Yongmin Kim[1]\**

[1]Department of Physics, Hanyang University, Seoul 31116, Korea

[2]Department of Physics, Dankook University, Cheonan 31116, Korea

**Corresponding Author**

\*E-mail: yongmin@dankook.ac.kr

Phone: +82-41-550-3422

Fax: +82-41-559-7859

ABSTRACT

Methylammonium-lead-halide compounds have emerged as promising bandgap engineering materials due to their ability to fine-tune the energy gap through halogen element mixing. We present a comprehensive investigation of the temperature-dependent photoluminescence (PL) transition characteristics exhibited by single crystals of chlorine and bromine-based methylammonium lead halides. $MAPbCl_3$ and $MAPbBr_3$ crystals exhibit a distinct sharp free exciton transition with an abrupt transition behavior associated with the structural phase transition as the temperature varies. However, when the two halogen elements are mixed within the crystals, no structural phase transition is observed. This study explores the temperature-dependent variations in integrated PL intensity, full-width-half-maximum, and peak transition energy of the crystals. The obtained results discuss the intricate interplay between temperature, crystal structure, and composition, providing valuable insights into the optical properties and potential applications of organic-inorganic hybrid methyl-ammonium lead halide single crystals as tunable energy gap semiconductor materials.

———

# Introduction

Recent advancements in perovskite materials have attracted considerable attention due to their exceptional solar efficiency. In particular, iodine-based perovskites have demonstrated impressive gains in photovoltaic devices by virtue of their near-infrared band gap energy. This has led to the integration of perovskite-silicon tandem solar cells, achieving a remarkable efficiency of over 33 %.[1] While efforts to enhance solar efficiency and improve the stability of lead-free perovskites remain prominent research goals, recent attention has also turned toward bromine and chlorine-based perovskites. These materials possess energy bands in the visible and near-ultraviolet regions, respectively, positioning them as highly promising candidates for LEDs[2-7] and photodetectors,[8-11] These advancements have opened up new possibilities for the development of high-performance, energy-efficient lighting, and sensing technologies. Moreover, significant progress has been made in replacing methylammonium (MA) with other organic materials such as formamidinium (FA) or butylammonium (BA), as well as substituting lead with tin. These alternative compositions exhibit improved stability and reduced toxicity concerns, further enhancing the practical applicability of perovskite materials. In addition, new research is being conducted from three-dimensional crystals to two-dimensional layered perovskite crystals. To fully exploit their capabilities in these applications, precise control over the band gap energy of perovskite materials is crucial. This control can be achieved through halogen element substitution or mixing.

Despite the extensive research conducted in these areas, there is still ample room for exploration regarding the physical properties of these materials. In this paper, we present our investigation on temperature-dependent photoluminescence (PL) transitions in various mixed halide perovskite single crystals. Through analysis of the PL spectra, we extract crucial information such as peak transition energy, intensity, and spectral linewidth (full width at half maximum; FWHM). These findings significantly contribute to our understanding of the fundamental physical characteristics of perovskite crystals. Furthermore, they shed light on the temperature-dependent behavior of these materials, providing valuable insights for their optimization and potential applications in optoelectronic devices. As a result, this research paves the way for further advancements and discoveries in the field of perovskite materials and their diverse applications in the realm of optoelectronics.

**Results and Discussion**

We performed an extensive analysis of various photoluminescence (PL) parameters related to the free exciton (FX) in $MAPbCl_xBr_{3-x}$ (x = 0 to 3 in 0.5-step) single crystals as a function of temperature. Figure 1 demonstrates the temperature-dependent photoluminescence (PL) transitions of selected samples of organic-inorganic hybrid methyl-ammonium lead halide single crystals. The PL spectra of the remaining samples are provided in the supplemental information (Fig. S1 and Fig. S2). In Figs. 1(a), (b), and (c), as well as Fig. S1 and Fig. S2, the red solid arrows represent the free exciton (FX) transition peaks. These FX peaks will be further analyzed and discussed in subsequent paragraphs, suggesting a comprehensive investigation of these important features. The insets of Figs. 1(a) and (b) reveal notable observations regarding the single-halide crystals, $MAPbCl_3$ and $MAPbBr_3$. These crystals exhibit a sudden spectral jump at approximately 169 K ($MAPbCl_3$) and 139 K ($MAPbBr_3$), indicating a well-known structural phase transition from a tetragonal to an orthorhombic crystal structure. This observation provides important insights into the crystal behavior and establishes a connection between temperature-induced crystal structure and the optical transition. In contrast, the mixed-halide crystal, $MAPbCl_{1.5}Br_{1.5}$., displays a broad and less distinct free exciton transition peak, as depicted in Fig. 1(c). This distinct behavior suggests different optical properties compared to the single-halide crystals. Similar transition behavior is observed in the spectra of other mixed crystal samples as seen in Figs. S1 and S2. These findings highlight the systematic investigation of various mixed-halide compositions, contributing to a comprehensive understanding of their optical characteristics. Furthermore, it has been established that all the mixed-halide samples do not exhibit a structural phase transition in the optical transition spectra as the temperature varies. This intriguing phenomenon has been attributed to the structural instability caused by the mixing of halogen elements. This result aligns with previous studies conducted by Galván et. al.,[12] who reported that the structure of mixed-halide crystals remains cubic throughout the entire temperature range by using temperature-dependent X-ray diffraction experiments, supporting and demonstrating the robustness and reliability of the current observations.

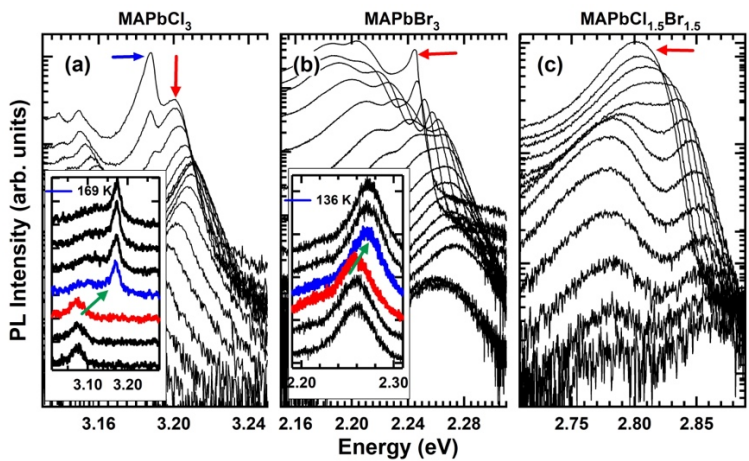

Figure 1. Temperature dependence of the PL transition spectra of (a) $MAPbCl_3$, (b) $MAPbBr_3$, and (c) $MAPbCl_{1.5}Br_{1.5}$ single crystals in steps of 10 K from 10 K to 150 K. Red arrows indicate FX transitions. The insets show the spectra for 1 K steps for the transition from tetragonal (red spectra) to orthorhombic (blue spectra) structure between 168 K and 169 K for $MAPbCl_3$ and between 135 K and 136 K for $MAPbBr_3$.

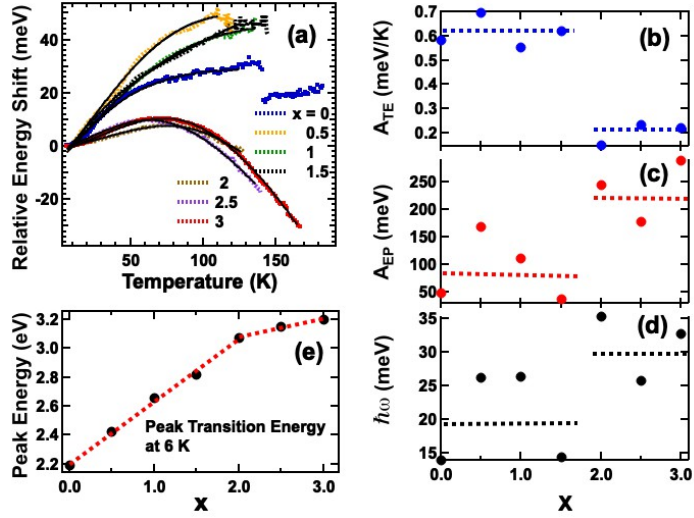

Figure 2. (a) Peak transition energies of the FX relative to those of at T = 6 K. Black solid lines are fitted values obtained by Eq. 1. (b) The coefficient related to the thermal contribution of the band gap. (c) The coefficient related to the electron-phonon interaction. (d) Phonon energy. (e) Peak transition energy at T = 6 K. Broken lines in (b) ~ (e) are guides for the eye.

In Fig. 2(a), we present the change in peak transition energy relative to the transition energy at 6 K. For $MAPbBr_3$, the estimated transition energy discontinuity as illustrated in Fig. 1(b) inset at 136 K is ~ 12.5 meV, which is attributed to the structural phase transition. However, in this figure, the phase transition depicted in Fig. 1(a) inset at 169 K for $MAPbCl_3$, the transition energy discontinuity ~ 88 meV) is too big to include in the figure. As seen in Fig. 2(a), in the case of x ≤ 1.5 we observe a gradual increase in the peak transition energies with temperature, which is a well-known characteristic of hybrid halide perovskite materials. However, a significant change in the trend of the band gap occurs when x > 1.5. In this case, the transition energy initially increases with temperature up to ~ 80 K and then decreases as the temperature further increases. This peculiar temperature dependence of the energy gap in semiconductors was described

TABLE I – Various parameters of MAPbCl$_x$Br$_{3-x}$ crystals obtained from the peak analysis of FX transitions

| parameters | x = 0 | 0.5 | 1 | 1.5 | 2 | 2.5 | 3 |
|---|---|---|---|---|---|---|---|
| $E$ (meV)* | 2.19 | 2.42 | 2.66 | 2.81 | 3.07 | 3.15 | 3.2 |
| $A_{TE}$ (meV/K) | 0.56 | 0.7 | 0.55 | 0.62 | 0.15 | 0.23 | 0.22 |
| $A_{EP}$ (meV) | 57.5 | 168.4 | 110.5 | 37.7 | 244.1 | 176.8 | 287.7 |
| $\hbar\omega$ (meV) | 15.45 | 26.23 | 26.4 | 14.27 | 35.35 | 25.74 | 32.81 |
| $E_{a1}$ (meV) | 25.01 | 74 | 85.1 | 75.89 | 88.5 | 20.79 | 44.39 |
| $E_{a2}$ (meV) | - | 15 | 8.5 | 6.54 | 21.3 | 8.97. | - |
| FWHM (meV)* | 3.18 | 15.46 | 12.87 | 14.39 | 8.76 | 7.22 | 5.71 |
| $\Gamma_0$ (meV) | 2.56 | - | 12.42 | 13.06 | 8.47 | 6.58 | 5.42 |
| $\gamma_{ac}$ (μeV/K) | 79 | - | 15 | 17 | - | 1.5 | 0.5 |
| $\gamma_{LO}$ (meV) | 11.7 | - | 4.57 | 14.1 | 5.73 ($\gamma_{imp}$) | 19.86 | 8.71 |
| $E_{LO}$ (meV) | 9.76 | - | 5.29 | 20.01 | 2.51 ($E_b$) | 19.83 | 10.24 |

*PL peak transition energy and FWHM at T = 6 K.*

phenomenologically by Varshni.[13] However, the Varshni equation does not provide a fundamental explanation for the temperature dependence of the semiconductor band gap. The combined effect of electron-phonon interaction and strong spin-orbit coupling is responsible for the temperature-dependent energy gap change in organic-inorganic hybrid lead halide crystals, which can be expressed as,[14-16]

$$E_g = E_0 + A_{TE}T - A_{EP}\left(\frac{2}{\exp(\hbar\omega/k_BT)+1}\right) \quad (1)$$

where $E_0$ is the band gap at $T = 0$ K, and $A_{TE}$ and $A_{EP}$ are the coefficients related to the thermal contribution of the band gap and electron-phonon interaction, respectively, and $\hbar\omega$ is the phonon energy. The black solid lines in Fig. 2(a) represent the results of the fitting, which align well with the equation. The parameters obtained from the equation are displayed in Fig. 2(b) ~ (d) and in Table 1. As observed in the figures, there is a trend that when x is less (more) than 1.5, the values of $A_{TE}$ are large (small), whereas the values of $A_{EP}$ and the phonon energy ($\hbar\omega$) exhibit the opposite behavior. It means that the phonon contribution is responsible for the decreasing band gap energy above 80 K for Cl-rich crystals (x ≥ 2). The peak transition energy at 6 K shows a linear increase in the range of 0 < x ≤ 1.5 , and then the slope changes above 2, as shown in Fig. 2(e). Specifically, the slope in the range of 0 < x ≤ 1.5 is 0.43 eV/x, while it is 0.13 eV/x for  2 ≤ x ≤ 3.

The temperature dependence of the integrated PL intensity is a valuable tool for investigating the behavior of bound states, such as excitons, in semiconductors. Figure 3 illustrates the integrated PL intensities as a function of temperature, normalized to the intensity at 6 K. In order to study the thermal dissociation of free excitons, we employed the one- or two-step Arrhenius equation[17] to fit the peak areal intensity for mixed-halide single crystals:

$$I = \frac{I_0}{1+C_1\exp{(-E_{a1}/k_BT)}+C_2\exp{(-E_{a2}/k_BT)}} \qquad (2)$$

where $I_0$ is the areal intensity at 0 K, $k_B$ is the Boltzmann constant, and $C_i$ and $E_{ai}$ ($i = 1, 2$) are the ratios to the thermal escape to the recombination rates and activation energies, respectively. The higher activation energy ($E_{a1}$) observed at elevated temperatures (> 35 K or equivalently, < 0.03 K$^{-1}$), as presented in Table 1 and Fig. 3(h), is associated with the thermal dissociation of excitons. This behavior reflects the binding energy of excitons formed within the crystals. Notably, the activation energies of the mixed halide crystals, except for the case when x=2.5 were found to be significantly higher than those of the unmixed samples. This discrepancy can be attributed to the random alloy characteristic of mixed halides. In mixed halide crystals, the positions of different halides are randomly distributed, resulting in a non-uniform fluctuation of the potential at the band edge. Consequently, the binding energy of an exciton bound to these random potentials is considerably greater than that of a free exciton in a flat-band case. The experimental data, summarized in Table 1 and illustrated in Figure 3 (a) for MAPbBr$_3$, were successfully fitted using a one-step Arrhenius function with $C_2 = 0$, indicating that the exciton binding energy of this crystal is comparable to other reported results. [18,19]

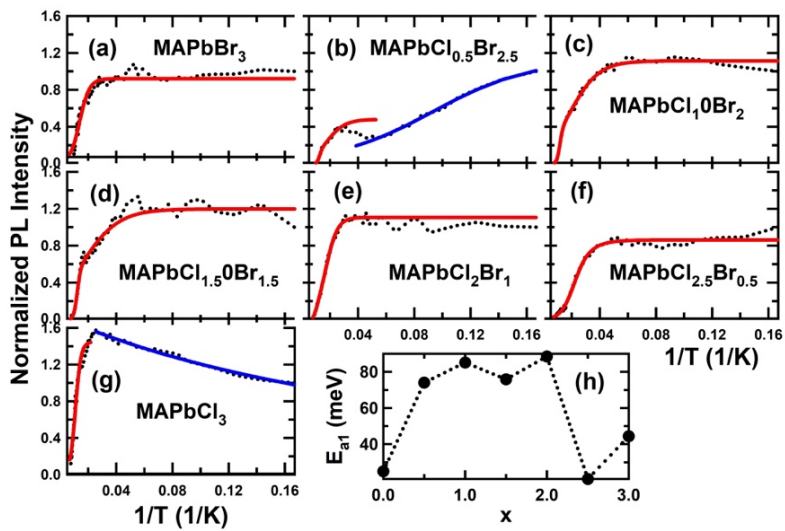

Figure 3. (a) ~ (g) Inverse temperature dependence of the integrated PL intensities. Broken lines are the integrated intensities and the solid lines are obtained from the Arrhenius fitting. (h) High temperature activation energy of samples obtained from the Arrhenius fitting.

As shown in Eq. 1, when the temperature is very low, the exponential terms converge to 0, and consequently, the intensity has to remain constant ($I_0$) at low temperatures. However, its intensity significantly increases or decreases with decreasing temperature for the case of MAPbCl$_{0.5}$Br$_{2.5}$ (Fig. 3(b)) and MAPbCl$_3$ (Fig. 3(g)), respectively. Unlike others, for the case of MAPbCl$_{0.5}$Br$_{2.5}$, as shown in Fig. S1 (a), the FX peak indicated by the red arrow is energetically too close to separating to a bound exciton (blue arrow). As a consequence, the increase in intensity at low temperature is likely to be attributed to the difficulty in separating the two peaks while fitting.

In the case of MAPbCl$_3$, because the distance between the two peaks marked by blue and red arrows is sufficiently far apart with narrow linewidths as seen in Fig. 1 (a), the intensity decrease at low temperature in the fitting of the FX peak (Fig. 3(g)) is reliable. As seen in Fig. 1 (a), when the low energy peak marked as a broken arrow starts to appear around 40 K, the peak intensity of FX (black solid arrow) begins to decrease. This reversal in intensity between the two peaks can be attributed to the kinetic energy transfer of carriers. When carriers undergo kinetic carrier transfer from high energy to low energy levels, the decrease in the FX transition peak with decreasing temperature can be expressed as the following equation, [20,21]

$$I = \frac{I_0}{1+\frac{K_0}{1+\exp(-\Delta E/k_B T)}} \quad (3)$$

where $K_0$ and $\Delta E$ are a constant and an energy difference between two adjacent levels, respectively. By fitting the experimental data, an obtained value of $\Delta E$ is ~ 19.7 meV, which is comparable to the 15.0 meV energy difference between the two adjacent transition peaks. This implies that carriers in the higher energy band (red arrow) undergo kinetic transfer to the lower energy band (blue arrow) as the temperature decreases, resulting in an intensity swap between the two peaks. In a previous report, we identified these two peaks as large (blue arrow) and small (red arrow) polarons by using PL transition experiments under strong magnetic fields to 60 T. [22]

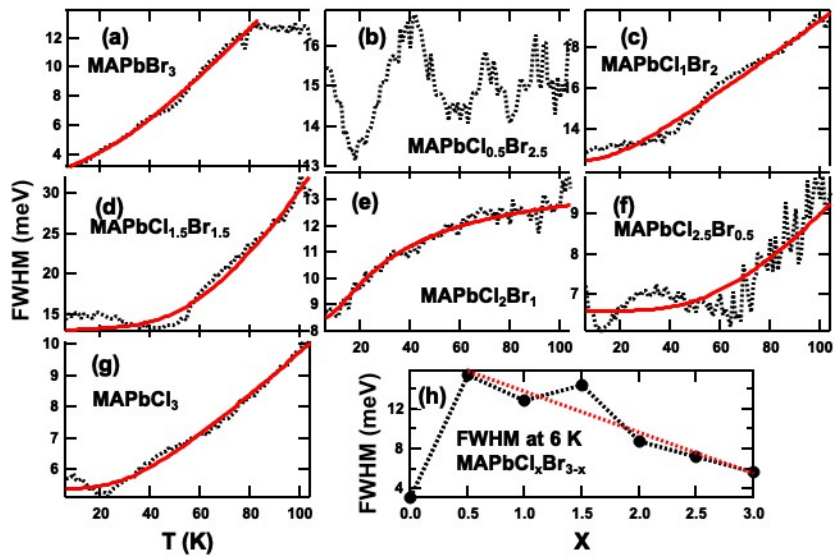

Figure 4 Figure 5. (a) ~ (g) Temperature dependence of the FWHM of PL transition spectra. Red solid lines are obtain by Eq. 4. (h) FWHM of samples at T = 6 K. Red dotted line is a guide for the eye.

The spectral linewidth (full width and half maximum; FWHM ($\Gamma$)) of an exciton transition peak provides valuable information about phonon scattering with carriers. The temperature dependence of the FWHM of PL spectra can be decomposed into three components: a temperature-independent linewidth $\Gamma_0$), contributions from acoustic phonons $\Gamma_{ac}$), and contributions from longitudinal optical (LO) phonons $\Gamma_{LO}$), which can be expressed as,[14,23,24]

$$\Gamma = \Gamma_0 + \Gamma_{ac} + \Gamma_{LO} = \Gamma_0 + \gamma_{ac}T + \frac{\gamma_{LO}}{\exp(E_{LO}/k_BT)}, \qquad (4)$$

where $\gamma_{ac}$ and $\gamma_{LO}$ are the coupling strength of the excitons with the acoustic and LO phonons, respectively, $E_{LO}$ is the LO phonon energy and $k_B$ denotes the Boltzmann constant. Upon examining Table 1, Fig. 4 (a) to (h), and Fig. S4, it becomes evident that the obtained values for different x-values lack consistency. Notably, the presence of a broad bound state transition in Fig. S1(a) for MAPbCl$_{0.5}$Br$_{2.5}$ undermines the reliability of the temperature dependence of the FWHM of the exciton (FX) (Fig. 4(b)). However, the value at 6 K in Fig. 4 (h) aligns with the temperature-independent $\Gamma_0$, as demonstrated by Fig. S4 (depicted by black triangle markers). This agreement arises because, at such low temperatures, phonon scattering is sufficiently suppressed, causing the FWHM to converge to a temperature-independent value. The larger $\Gamma_0$ value observed for the mixed halide crystal is consistent with the structural instability of this phase, as revealed by XRD experiments. The obtained $\gamma_{ac}$ value (79 µeV/K) for MAPbBr$_3$ is comparable to the reference value ~ 70 µeV/K). Additionally, $\gamma_{ac}$ tends to decrease as x increases. This trend can be attributed to a decrease in acoustic phonon interaction with an increasing amount of chlorine (Cl) in the PX6 cage.

As seen in Fig. 3(b) and Fig. 4(b), for the case of MAPbCl$_{0.5}$Br$_{2.5}$, we could not obtain a consistent way of fitting results for both PL peak intensity and FWHM, respectively. This is possibly due to a big side peak marked as a red arrow which is located energetically close proximity to the main transition. In Fig. 4 (e), the MAPbCl$_2$Br$_1$ sample shows different behavior in comparison to the other samples. The amount of the FWHM saturates with increasing temperature. This kind of behavior is due to the interaction of the phonon with the impurities $\Gamma_{imp} = \gamma_{imp}\exp(-E_b/k_BT)$ with an average binding energy $E_b$.[23,25] It is challenging to provide a clear explanation as to why impurity scattering has such a significant effect only in this particular sample.

**Conclusion**

In conclusion, we measured the PL transitions of various organic-inorganic hybrid perovskite single crystals as a function of temperature. The single halide crystals show sharp exciton transition features with structural phase transition. Because of the phase instability, the mixed halide crystals show relatively broad exciton PL transitions without phase transition. Peak transition energy analysis revealed that there are two types of transition energy behavior. When x ≤ 1.5, the peak transition energy increases monotonically with increasing the temperature from 6

K, whereas when x > 1.5, the peak transition energy turns over to diminishing the transition energy above ~ 80 K. These temperature-dependent band gap behavior is associated with the thermal expansion of lattice and the electron-phonon interaction. $A_{TE}$ increases the energy gap with increasing temperature, whereas $A_{EP}$ and $\hbar\omega$ do the opposite. In the case of Br-rich samples, the energy gap increases with increasing temperature because of the large $A_{TE}$ and small $A_{EP}$, whereas in the case of Cl-rich samples, these values are the opposite, so the energy gap decreases at temperatures above 80 K.

By presenting these findings, we demonstrate the influence of temperature on the integrated PL intensity and FWHM, which provide insights into the thermal dissociation of excitons and the phonon scattering induced line broadening, respectively, in organic-inorganic hybrid methyl-ammonium lead halide single crystals. Our results highlight the enhanced activation energies observed in mixed halide crystals, which can be attributed to their random alloy nature and the associated fluctuations in the potential landscape at the band edge.

Overall, our findings contribute to a comprehensive understanding of the optical properties and temperature-dependent behavior of organic-inorganic hybrid perovskite single crystals. The observed distinctions between single halide and mixed halide crystals, as well as the different transition energy behaviors, provide valuable insights for the design and optimization of these materials for various applications in optoelectronic devices and related fields.

**Experimental Section**

All chemicals employed to grow samples were used as received without further purification. Methylammonium halides ($CH_3NH_3X$; X = Cl and Br) were sourced from Greatcell Solar Materials. Lead (II) halides (X = Cl and Br) were purchased from Alfa Aesar. Reagent-grade organic solvents such as γ-butyrolactone (GBL), dimethyl sulfoxide (DMSO), and dimethyl formamide (DMF) were purchased from Aldrich. $MAPbX_3$ (MA = $CH_3NH^+_3$ (methylammonium), X = $Cl^-$ and $Br^-$) single crystals were prepared using the inverse temperature crystallization (ITC) method.[26,27] Details of the sample fabrication methods used for this study can be found elsewhere.[28]

The temperature-dependent PL measurements were conducted using a double-stage helium gas closed cycle refrigerator, which allowed precise control of the sample temperature ranging from 6 K to room temperature. To ensure accurate temperature readings of the sample mounted on the refrigerator's cold finger, two thermometers were strategically positioned on the top and bottom sides of the samples. Photoluminescence (PL) spectra were recorded at each 1 K-step increment. For the PL measurements, the excitation source was carefully selected based on the emission range of each sample. Three different laser lines were used: 325 nm and 442 nm from a HeCd laser, and 532 nm from an Nd:YAG laser. A 50-cm spectrograph equipped with a 600 line/mm grating and a liquid nitrogen-cooled charge-coupled device (CCD) with a pixel size of 14 μm (2048 by 512 pixel) was employed to detect the PL signal. The spectrographic system exhibited a spectral resolution of ~ 0.6 meV at 540 nm.

ACKNOWLEDGMENT

The present research was supported by the research fund of Dankook University in 2022.

ACKNOWLEDGMENT

The present research was supported by the research fund of Dankook University in 2022.

REFERENCES

(1) National Renewable Energy Laboratory Best Research-Cell Efficiency Chart. https://www.nrel.gov/pv/assets/pdfs/best-research-cell-efficiencies.pdf (accessed on February 2023).

(2) Vashishtha, P.; Halpert J. E. Field-Driven Ion Migration and Color Instability in Red-Emitting Mixed Halide Perovskite Nanocrystal Light-Emitting Diodes. *Chem. Mater*. **2017**, 29, 5965 - 5973.

(3) Lin, K.; Xing, J.; Quan, L. N.; de Arquer, F. P. G.; Gong, X.; Lu, J.; Xie, L.; Zhao, W.; Zhang, D.; Yan, C.; Li, W.; Liu, X.; Lu, Y.; Kirman, J.; Sargent, E. H.; Xiong, Q.; Wei, Z. Perovskite light-emitting diodes with external quantum efficiency exceeding 20 per cent. *Nature* **2018**, 562, 245 - 249.

(4) Lee, D.-K.; Shin, Y.; Jang, H. J.; Lee, J.-H.; Park, K.; Lee, W.; Yoo, S.; Lee, J. Y.; Kim, D.; Lee, J.-W.; Park N.-G. Nanocrystalline Polymorphic Energy Funnels for Efficient and Stable Perovskite Light Emitting Diodes. *ACS Energy Lett*. **2021**, 6, 1821 - 1830.

(5) Zhang, K.; Zhu, N.; Zhang, M.; Wang, L.; Xing, J. Opportunities and challenges in perovskite LED commercialization. *J. Mater. Chem. C* **2021**, 9, 3795 -3799.

(6) Dyrvik, E. G.; Warby, J. H.; McCarthy, M. M.; Ramadan, A. J.; Zaininger, K.-A.; Lauritzen, A. E.; Mahesh, S.; Taylor, R. A.; Snaith, H. J. Reducing Nonradiative Losses in Perovskite LEDs through Atomic Layer Deposition of $Al_2O_3$ on the Hole-Injection Contact. *ACS Nano* **2023**, 17, 3289 - 3300.

(7) Jang, G.; Han, H.; Ma, S.; Lee, J.; Lee, C. U.; Jeong, W.; Son, J.; Cho, D.; Kim, J.-H.; Park, C.; Moon, J. Rapid crystallization-driven high-efficiency phase-pure deep-blue Ruddlesden–Popper perovskite light-emitting diodes. *Adv. Photonics* **2023**, 5(1), 016001-1 - 016001-11.

(8) Rybin, N.; Ghosh, D.; Tisdale, J.; Shrestha, S.; Yoho, M.; Vo, D.; Even, J.; Katan, C.; Nie, W.; Neukirch, A. J.; Tretiak, S. Effects of Chlorine Mixing on Optoelectronics, Ion Migration, and Gamma-Ray Detection in Bromide Perovskites. *Chem. Mater.* **2020**, 32, 1854 -1863.

(9) Li, C.; Ma, Y.; Xiao, Y.; Shen, L.; Ding, L. Advances in perovskite photodetectors. *InfoMat* **2020**, 2, 1247 - 1256.

(10) Liu, Y.; Zhang, Y.; Zhu, X.; Feng, J.; Spanopoulos, I.; Ke, W.; He, Y.; Ren, X.; Yang, Z.; Xiao, F.; Zhao, K.; Kanatzidis, M.; Liu, S. (F). Triple-Cation and Mixed-Halide Perovskite Single Crystal for High-Performance X-ray Imaging. *Adv. Mater.* **2021**, 33, 2006010.

(11) Wang, H.; Sun, Y.; Chen, J.; Wang, F.; Han, R.; Zhang, C.; Kong, J.; Li, L.; Yang, J. *Nanomaterials* **2022**, 12, 4390.

(12) Alvarez-Galván, C.; Alonso, J. A.; López, C. A.; L\ópez-Linares, E.; Contreras, C.; Lázaro, M. J.; Fauth, F.; Martínez-Huerta, M. V. Crystal Growth, Structural Phase Transitions, and Optical Gap Evolution of $CH_3NH_3Pb(Br_{1-x}Cl_x)_3$ Perovskites. *Cryst. Growth Des*. **2019**, 19, 918 - 924.

(13) Varshni, Y. P. Temperature dependence of the energy gap in semiconductors. *Physica* **1967**, 34, 149 - 154.

(14) Bhosale, J.; Ramdas, A. K.; Burger, A.; Muñoz, A.; Romero, A. H.; Cardona, M.; Lauck, R.; Kremer, R. K. Temperature dependence of band gaps in semiconductors: Electron-phonon interaction. *Phys. Rev. B* **2012**, 86, 195208-1 - 195208-10.

(15) Wang, S.; Ma, J.; Li, W.; Wang, J.; Wang, H.; Shen, H.; Li, J.; Wang, J.; Luo, H.; Li, D. Temperature-Dependent Band Gap in Two-Dimensional Perovskites: Thermal Expansion

Interaction and Electron-Phonon Interaction. J. Phys. Chem. Lett. 2019, 10, 2546 - 2553.

(16) Yu, S.; Xu, J.; Shang, X.; Ma, E.; Lin, F.; Zheng, W.; Tu, D.; Li, R.; Chen, X. Unusual Temperature Dependence of Bandgap in 2D Inorganic Lead-Halide Perovskite Nanoplatelets. *Adv. Sci.* **2021**, 8, 2100084.

(17) Lu, T.; Ma, Z.; Du, C.; Fang, Y.; Wu, H.; Jiang, Y.; Wang, L.; Dai, L.; Jia, H.; Liu, W.; Chen, H. Temperature-dependent photoluminescence in light-emitting diodes. *Sci. Rep.* **2014**, 4, 6131.

(18) Lozhkina, O. A.; Yudin, V. I.; Murashkina, A. A.; Shilovskikh, V. V.; Davydov, V. G.; Kevorkyants, R.; Emeline, A. V.; Kapitonov, Y. V.; Bahnemann, D. W. Low Inhomogeneous Broadening of Excitonic Resonance in MAPbBr3 Single Crystals. *J. Phys. Chem. Lett.* **2018**, 9, 302 - 305.

(19) Tilchin, J.; Dirin, D. N.; Maikov, G. I.; Sashchiuk, A.; Kovalenko, M. V.; Lifshitz, E. Hydrogen-like Wannier–Mott Excitons in Single Crystal of Methylammonium Lead Bromide Perovskite. *ACS Nano* **2016**, 10, 6363 - 6371.

(20) Popescu, D. P.; Eliseev, P. G.; Stintz, A.; Malloy, K. J. Temperature dependence of the photoluminescence emission from InAs quantum dots in a strained $Ga_{0.85}In_{0.15}As$ quantum well. *Semicond. Sci. Technol.* **2004**, 19, 33 - 38.

(21) Shin, Y. H.; Kim, Y.; Song, J. D. Optically detected kinetic carrier transfer in InP-GaP lateral nanowires. *J. Lum.* **2018**, 202, 107 - 110.

(22) Shin, Y. H.; Choi, H.; Park, C.; Park, D.; Jeong, M. S.; Nojiri, H.; Yang, Z.; Kohama, Y.; and Kim, Y. Combination of optical transitions of polarons with Rashba effect in methylammonium lead trihalide perovskites under high magnetic fields. *Phys. Rev. B* **2021**, 104, 035205.

(23) Lee, J.; Koteles, E. S.; Vassell, M. O. Luminescence linewidths of excitons in GaAs quantum wells below 150 K. *Phys. Rev. B* **1986**, 33, 5512 - 5516.

(24) Rudin, S.; Reinecke, T. L.; Segall, B. *Phys. Rev. B* **1990**, 42, 11218–11231.

(25) Wright, A. D.; Verdi, C.; Milot, R. L.; Eperon, G. E.; Pérez-Osorio, M. A.; Snaith, H. J.; Giustino, F.; Johnston, M. B.; Herz, L. M. Electron–phonon coupling in hybrid lead halide perovskites. *Nat. Commun.* **2016**, 7, 11755.

(26) Maculan, G.; Sheikh, A. D.; Abdelhady, A. L.; Saidaminov, M. I.; Haque, M. A.; Murali, B.; Alarousu, E.; Mohammed, O. F.; Wu, T.; Bakr, O. M. $CH_3NH_3PbCl_3$ Single Crystals: Inverse Temperature Crystallization and Visible-Blind UV-Photodetector. *J. Phys. Chem. Lett.* **2015**, 6, 3781 - 3786.

(27) Saidaminov, M. I.; Abdelhady, A. L.; Murali, B.; Alarousu, E.; Burlakov, V. M.; Peng, W.; Dursun, I.; Wang, L.; He, Y.; Maculan, G.; Goriely, A.; Wu, T.; Mohammed, O. F.; and Bakr, O. M. High-quality bulk hybrid perovskite single crystals within minutes by inverse temperature crystallization. *Nature Commun.* **2015**, 6, 7856.

(28) Park, D.Y.; Byun, H.R.; Lee, A.Y.; Choi, H. M.; Lim, S. C.; Jeong, M. S. Synthesis and Characterization of Bandgap-modulated Organic Lead Halide Single Crystals. *J. Kor. Phys. Soc.* **2018**, 73, 1716–1724.

# Temperature Dependence of the Optical Transition Characteristics of MAPbCl$_x$Br$_{3-x}$ Single Crystals

D. Y. Park[1], Y. H. Shin[2], and Yongmin Kim[2]*

[1]Department of Physics, Hanyang University, Seoul 04763, Korea and

[2]Department of Physics, Dankook University, Cheonan 31116, Korea (Dated: July 12, 2023)

SUPPLEMENTARY INFORMATION

Figure 1 in the main text, the temperature dependent photoluminescence (PL) spectra of three selected samples among seven sample were displayed. The rest of four sample spectra are showed in Figure S1 (MAPbCl$_{0.5}$Br$_{2.5}$ and MAPbCl$_1$Br$_2$) and Figure S2 (MAPbCl$_1$Br$_1$ and MAPbCl$_{2.5}$Br$_{0.5}$). As seen in Figure S1(a) for MAPbCl$_{0.5}$Br$_{2.5}$, a strong transition peak (blue arrow) is located close to the low energy side of the main transition peak (red arrow). This peak made it difficult to separate the main transition peak in multi-oscillator fitting, which made analysis difficult. As a consequence, as seen in Figures 3(b) and 4(b) in the main text, the peak intensity and FWHM of this sample in varying temperature made it difficult to analyze unlike other samples.

Figure S3 shows the PL peak transition energy measured at 6 K for our sample (blue open squares), and the optical band gap (solid blue square) along with lattice constant (solid red circle) extracted from Ref. 12 in the main text. The optical transition energy difference comes from the low and room temperature measurement for ours and Ref. 12, respectively.

Figure S4 exhibits various parameters obtained from the FWHM fitting in Figure 4 in the main text. LO phonon related parameters (blue open and filled squares) do not show any relevant relationship with respect to x-values. Acoustic phonon parameter $\gamma_{ac}$ decreases with decreasing x, except in the case of x = 2.0. Temperature-independent parameter $\Gamma_0$ (solid black triangle) is the smallest in the case of x = 0 and 3, and decreases with decreasing x-values between 0 and 3 in the case of mixed halide samples.

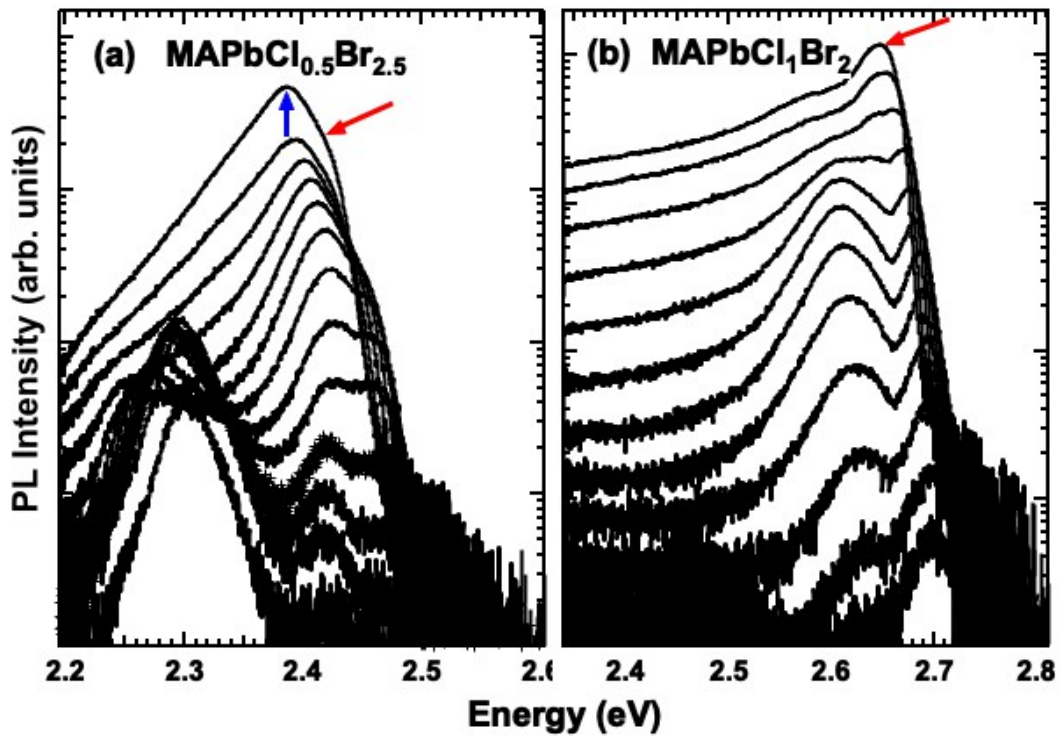

**Figure S 1.** Temperature dependence of the PL transition spectra of (a) MAPbCl$_{0.5}$Br$_{2.5}$ and (b) MAPbCl$_1$Br$_2$ single crystals from 6 K to 126 K in steps of 10 K. The red arrow indicates FX transition. The peak indicated by the blue arrow in (a) is located close to the FX peak (red arrow), which makes it difficult to analyze the FX transition.

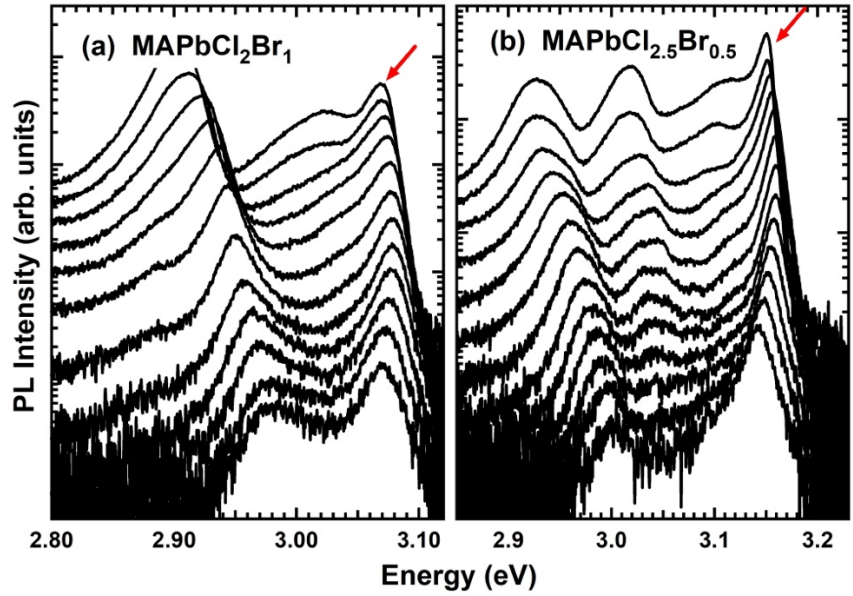

**Figure S 2.** Temperature dependence of the PL transition spectra of (a) MAPbCl$_2$Br$_1$ and (b) MAPbCl$_{2.5}$Br$_{0.5}$ single crystals from 6 K to 126 K in steps of 10 K. Red arrow indicates FX transition.

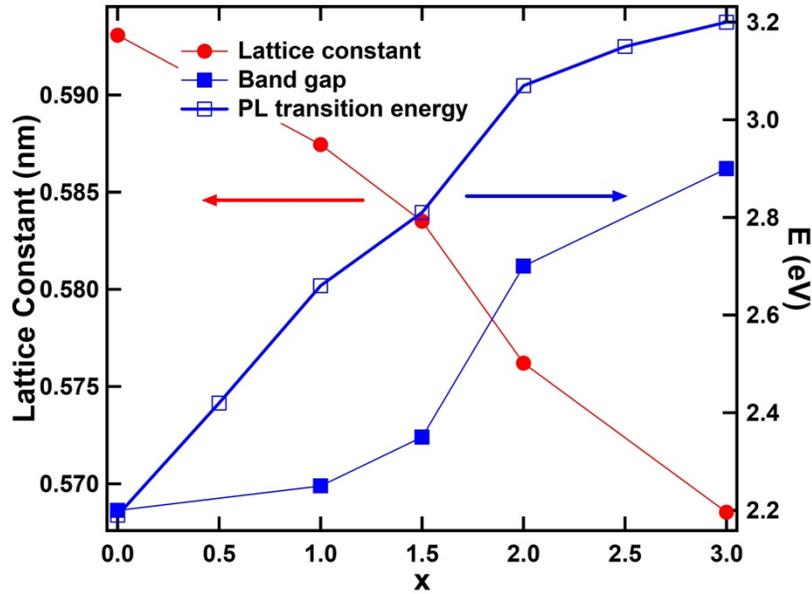

**Figure S 3.** Room temperature lattice constant (left axis), optical band gap and PL peak transition energy (right axis) with respect to x. Lattice constant and band gap energy are extracted from Ref. 12 in the main text. PL peak transition energies (blue open square markers) are measured at T = 6 K (Fig. 2(b) in the main text).

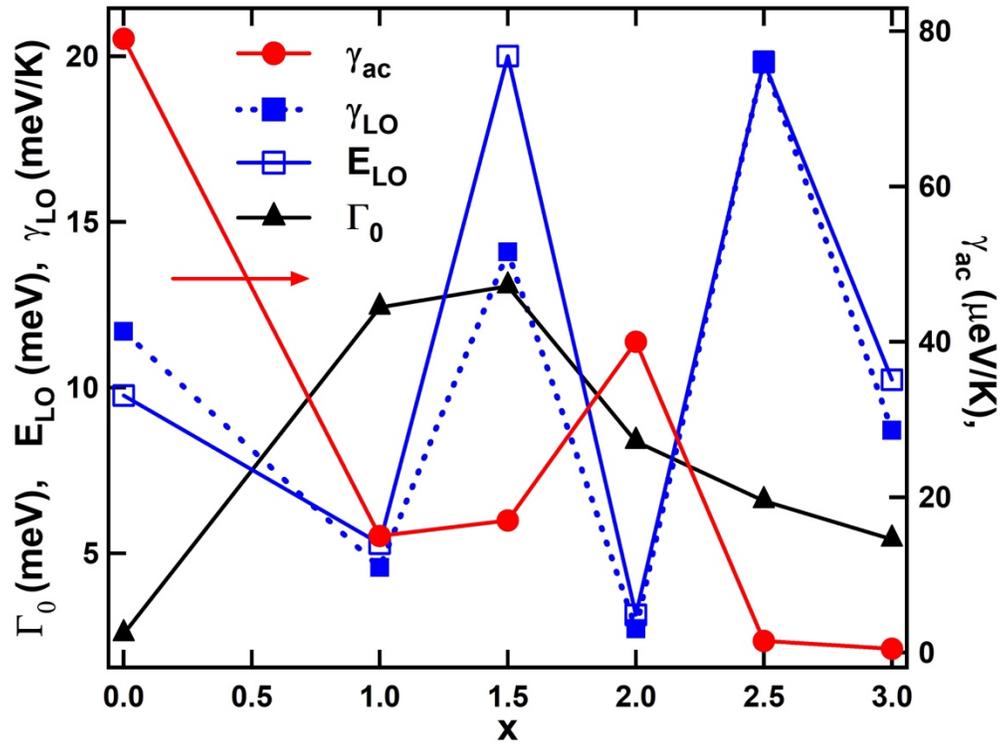

**Figure S 4.** Various parameters obtained from the FWHM fitting in Fig. 4 in the main text.